\documentclass[12pt]{article}
\usepackage{graphicx}

\newcommand\pubnumber{TTK-22-38}
\newcommand\pubdate{\today}

\def\institute{Institute for Theoretical Particle Physics and Cosmology, RWTH Aachen University\\
	Sommerfeldstrasse 16, D-52056 Aachen, Germany}

\def\Title#1{\begin{center} {\Large #1 } \end{center}}
\def\Author#1{\begin{center}{ \sc #1} \end{center}}
\def\Address#1{\begin{center}{ \it #1} \end{center}}

\newcommand\pubblock{\rightline{\begin{tabular}{l} \pubnumber\\
         \pubdate  \end{tabular}}}
\newenvironment{Abstract}{\begin{quotation}  }{\end{quotation}}
\newenvironment{Presented}{\begin{quotation} \begin{center} 
             PRESENTED AT\end{center}\bigskip 
      \begin{center}\begin{large}}{\end{large}\end{center} \end{quotation}}
\def\Acknowledgements{\bigskip  \bigskip \begin{center} \begin{large}
             \bf ACKNOWLEDGEMENTS \end{large}\end{center}}

\def\beq{\begin{equation}}
\def\eeq#1{\label{#1}\end{equation}}
\def\eeqn{\end{equation}}

\def\beqa{\begin{eqnarray}}
\def\eeqa#1{\label{#1}\end{eqnarray}}
\def\eeqan{\end{eqnarray}}

\let\bar=\overbar

\def\Dslash{\not{\hbox{\kern-4pt $D$}}}
\def\dslash{\not{\hbox{\kern-2pt $\del$}}}

\def\msb{{\bar{\ssstyle M \kern -1pt S}}}

\begin{document}
\begin{titlepage}
\pubblock

\vfill
\Title{Top-pair events with B-hadrons at the LHC}
\vfill
\Author{Terry Generet}
\Address{\institute}
\vfill
\begin{Abstract}
The first implementation of fragmentation in a numerical code for the computation of cross sections at next-to-next-to-leading order in QCD has recently been completed. I will present some results of the first application of this new framework to the production of top-quark pairs at the LHC in association with a bottom-flavoured hadron. Additionally, I will present an extended version of this calculation, which includes the decay of the $B$-hadron to a $J/\psi$ meson or a muon.
\end{Abstract}
\vfill
\begin{Presented}
$15^\mathrm{th}$ International Workshop on Top Quark Physics\\
Durham, UK, 4--9 September, 2022
\end{Presented}
\vfill
\end{titlepage}
\def\thefootnote{\fnsymbol{footnote}}
\setcounter{footnote}{0}

\section{Introduction}
Over the years, many different methods of measuring the top-quark mass have been proposed. Many of these methods rely on observables involving $b$-quarks from top-quark decays, but one particularly promising class of measurements substitutes $b$-quarks for $B$-hadrons. This substitution can significantly reduce the systematic uncertainty of the measurement. Consequently, this type of final state has been studied in the context of top-quark-mass measurements many times in the past \cite{Kharchilava:1999yj,Biswas:2010sa,Agashe:2012bn,Agashe:2016bok}.

However, many of these studies were performed using parton showers, an approach which was found in ref.~\cite{Biswas:2010sa} to limit the precision of the extracted top-quark mass to a few GeV. The alternative is to use fixed-order calculations, where the theoretical uncertainties can be systematically controlled. Past studies were limited to next-to-leading order (NLO) calculations, such as the one presented in ref.~\cite{Biswas:2010sa}. That study concluded that using NLO fixed-order calculations improves the theoretical limit on the precision of the top-quark mass determination to roughly $1$ GeV. While this was a significant improvement over what could be achieved using parton showers, it is nevertheless not precise enough to compete with modern sub-GeV-precision top-quark-mass measurements.

To improve on this situation, a next-to-next-to-leading order (NNLO) calculation is needed. Ref.~\cite{Czakon:2021ohs} presented the first implementation of fragmentation in a general code for NNLO computations \cite{Czakon:2010td,Czakon:2011ve,Czakon:2014oma,Czakon:2019tmo}. While the implementation is fully general and can be used to study any process involving any final-state hadron, its first application was to top-quark-pair production at the LHC in association with a $B$-hadron \cite{Czakon:2021ohs}. These results were later improved upon in ref.~\cite{Czakon:2022pyz} by fitting a new set of $B$-hadron fragmentation functions and by incorporating the decay of the $B$-hadron to a $J/\psi$ meson or a muon. I will now present some of the results of refs.~\cite{Czakon:2021ohs,Czakon:2022pyz}.

\section{Results}
One of the observables studied in ref.~\cite{Czakon:2021ohs} is of particular relevance to future fits of $B$-hadron fragmentation functions: the ratio of the transverse momentum of the $B$-hadron to the transverse momentum of the jet that contains it, i.e.~$p_T(B)/p_T(j_B)$. This observable is a very close proxy for the fragmentation function. It is expected to be highly sensitive to the shape of the fragmentation function, while being insensitive to the shape of the parton distribution functions (PDFs). Sensitivity to the PDFs has always been a concern when fitting fragmentation functions to hadron-collider data, and is one of the reasons why $B$-hadron fragmentation function fits typically rely on $e^+e^-$-collider data only \cite{Corcella:2001hz,Cacciari:2002re,Corcella:2005dk,Cacciari:2005uk,Kniehl:2007erq,Fickinger:2016rfd}. The new fit presented in ref.~\cite{Czakon:2022pyz} follows this trend, being based on data from the ALEPH \cite{ALEPH:2001pfo}, DELPHI \cite{DELPHI:2011aa}, OPAL \cite{OPAL:2002plk} and SLD \cite{SLD:2002poq} collaborations.

\begin{figure}[t]
	\includegraphics[width=0.49\textwidth]{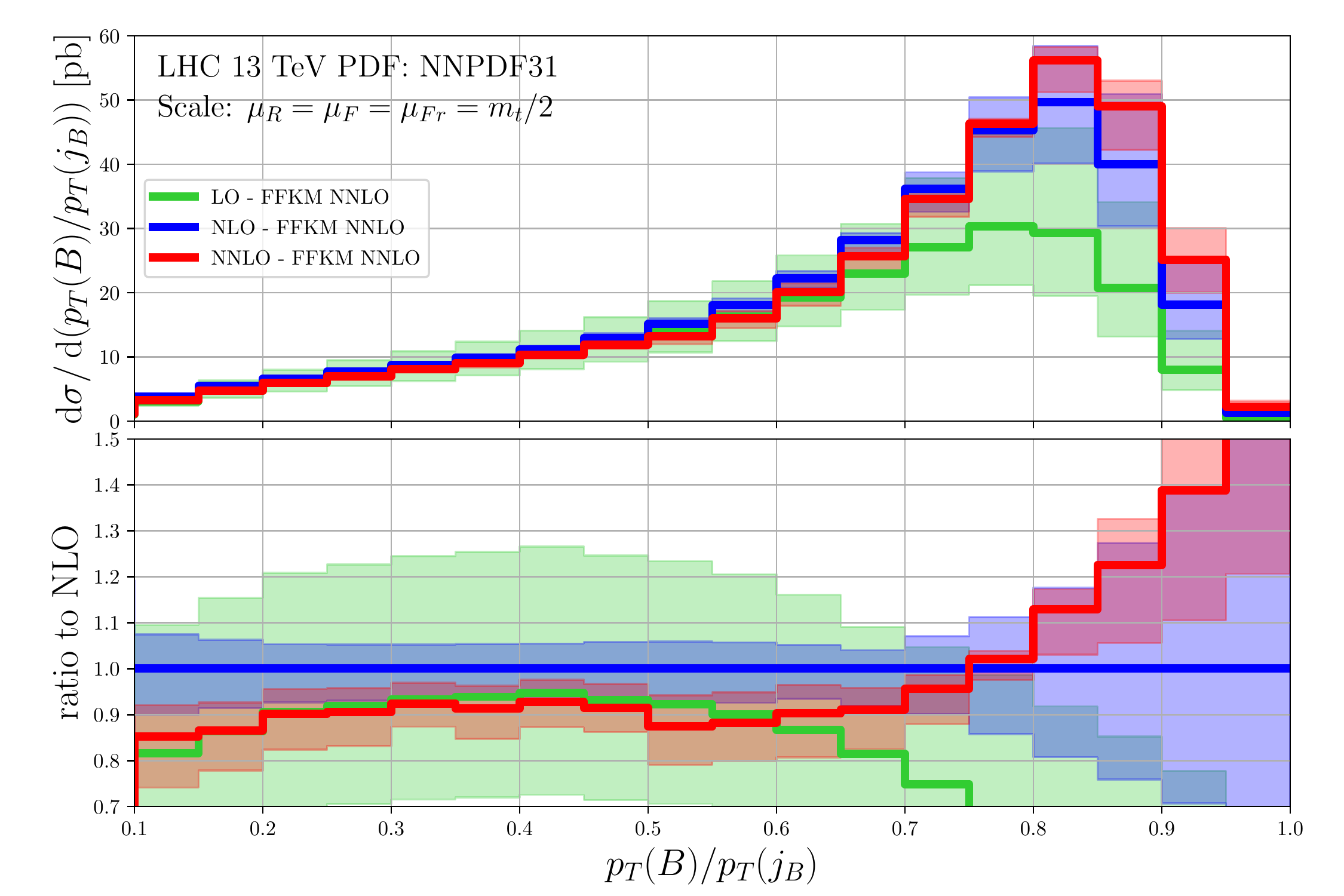}%
	\includegraphics[width=0.49\textwidth]{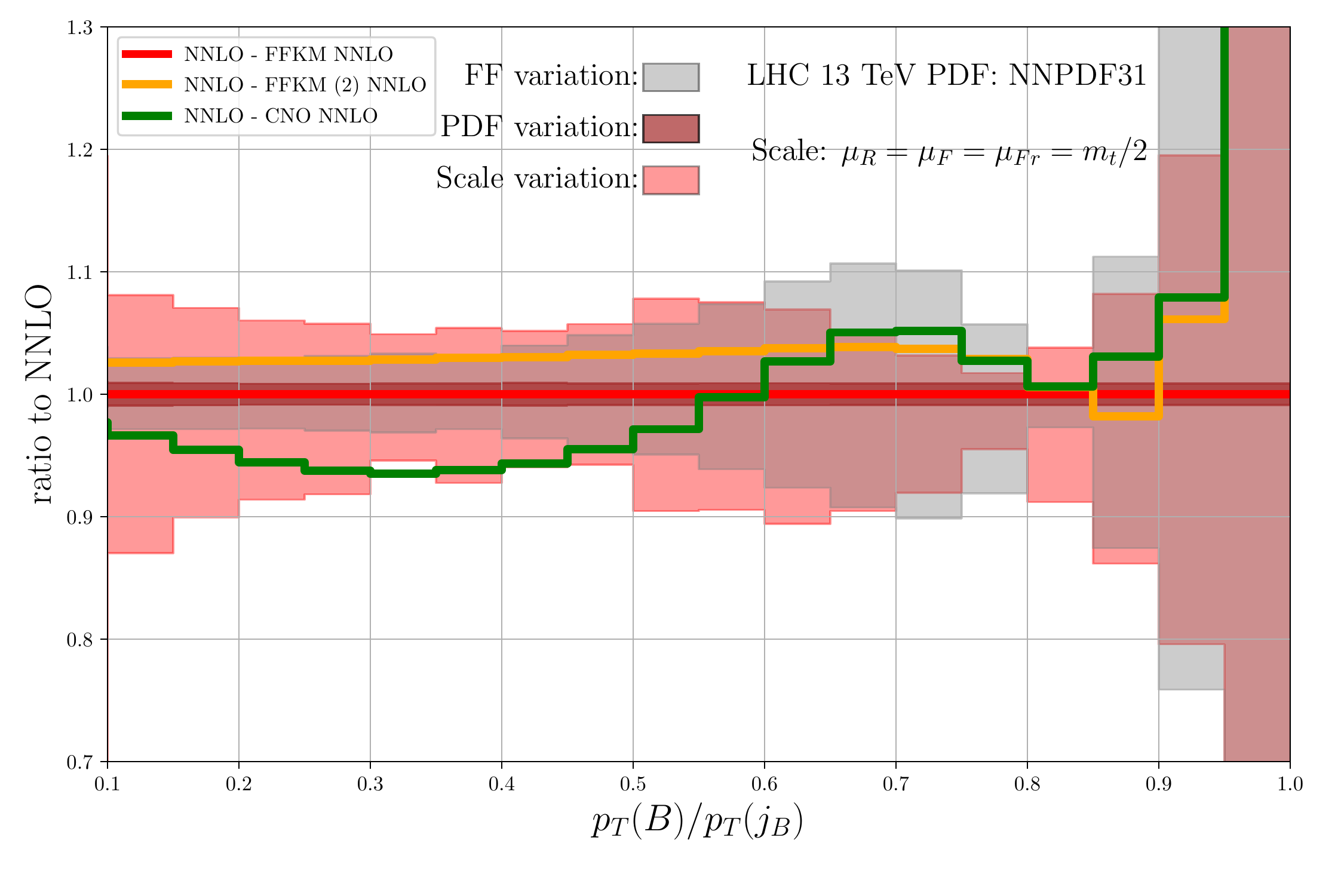}
	\caption{Left: the distribution of $p_T(B)/p_T(J_B)$, the ratio of the transverse momenta of the $B$-hadron and the jet that contains it, at leading order (green), NLO (blue) and NNLO (red). Right: a comparison of the sizes of different uncertainties.}
	\label{fig:jetRatio}
\end{figure}

Fig.~\ref{fig:jetRatio} shows the distribution of $p_T(B)/p_T(j_B)$ obtained in ref.~\cite{Czakon:2021ohs}. Looking at the left panel, increasing the perturbative order from leading order (LO) to NLO significantly reduces the scale uncertainties. Including the NNLO corrections reduces the scale uncertainties by another factor of two around the peak. This highlights the importance of the NNLO corrections to high-precision fragmentation function fits. The right panel compares different sources of theory uncertainties. The fragmentation function uncertainty band (grey) was found to be much wider for this observable than for the others studied in ref.~\cite{Czakon:2021ohs}, proving that $p_T(B)/p_T(j_B)$ is indeed unusually sensitive to the fragmentation function. Additionally, the PDF uncertainty band (dark red) is tiny and almost completely flat, showing that this observable is indeed PDF-insensitive. Combined, these two properties imply that, in the future, $B$-hadron fragmentation functions could be fitted to LHC data using a similar observable.

In the context of top-quark-mass measurements, one of the most commonly studied observables is the invariant mass of the $B$-hadron or one of its decay products with the charged lepton from the top-quark decay. Refs.~\cite{Czakon:2021ohs,Czakon:2022pyz} studied top-pair production in the dileptonic channel, i.e. there are two charged leptons in the final state. Since determining the charge of the $B$-hadron can be difficult experimentally, the observable should be blind to the $B$-hadron charge. This leads to two possible choices of the invariant mass for each event. Ideally, the $B$-hadron and the lepton would always belong to the same top-quark decay. While this is impossible to achieve, a good approximation is to always pick the smaller of the two possible invariant masses. This observable is called $m(B\ell)_{\rm min}$.

\begin{figure}[t]
\includegraphics[width=0.32\textwidth]{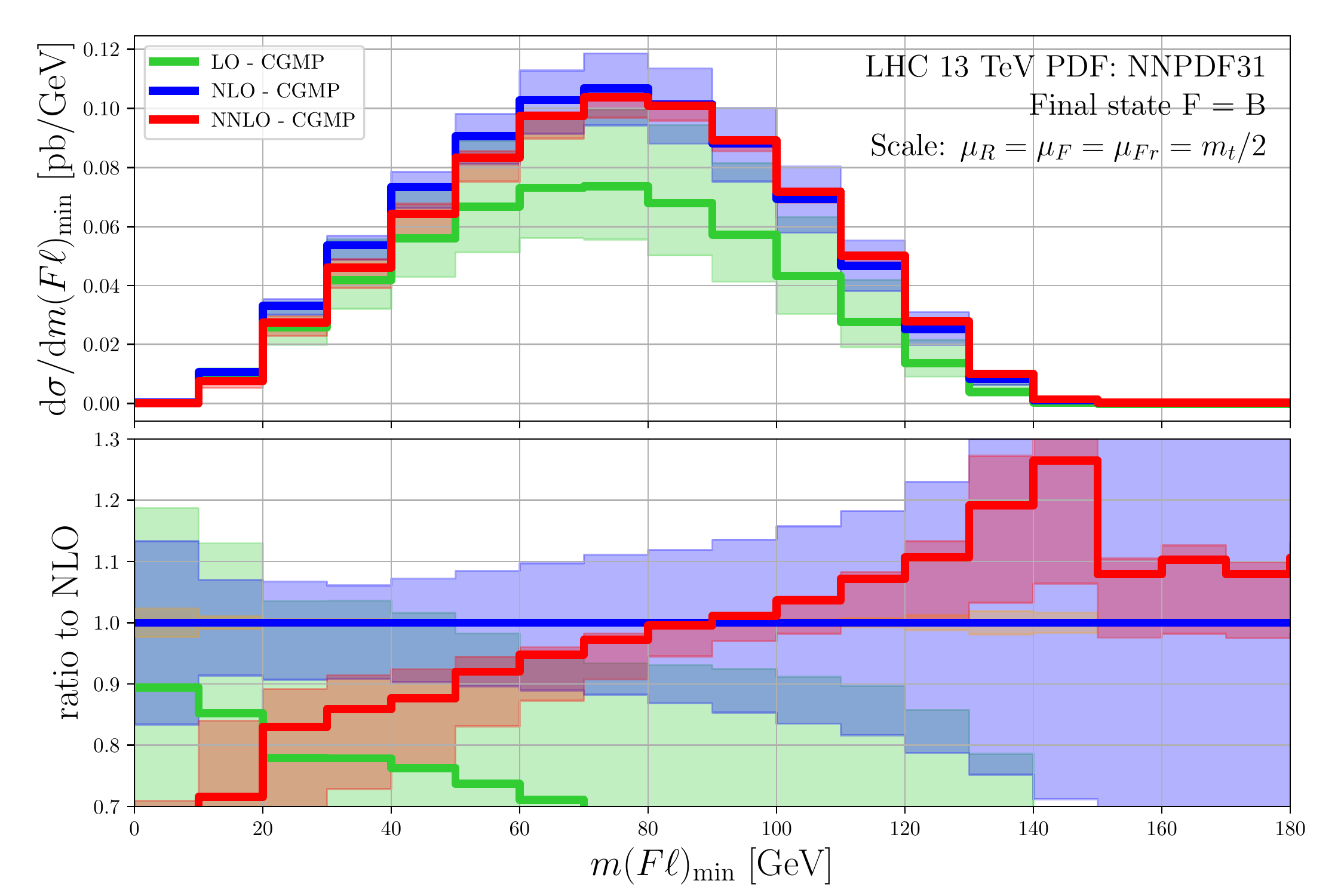}
\includegraphics[width=0.32\textwidth]{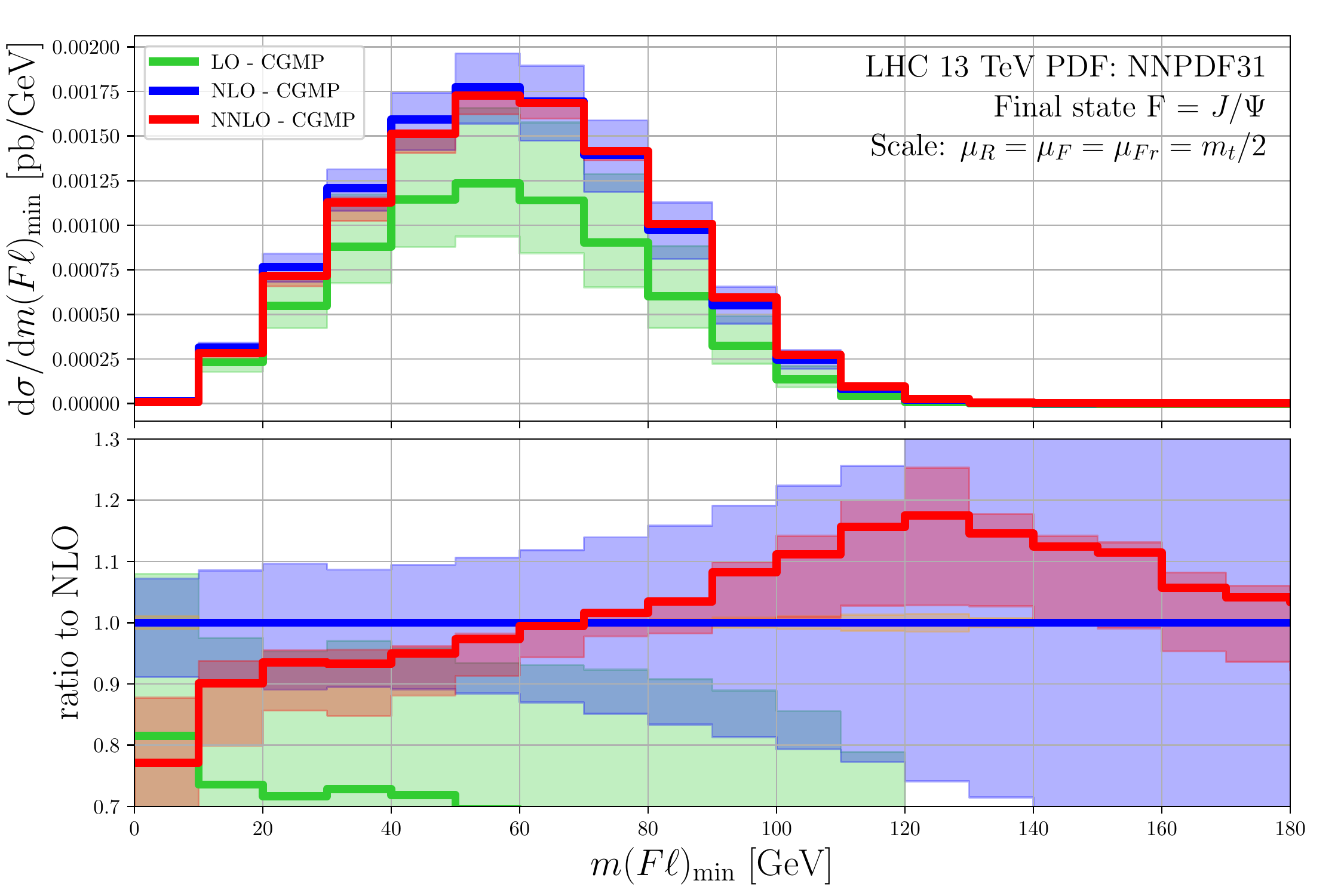}
\includegraphics[width=0.32\textwidth]{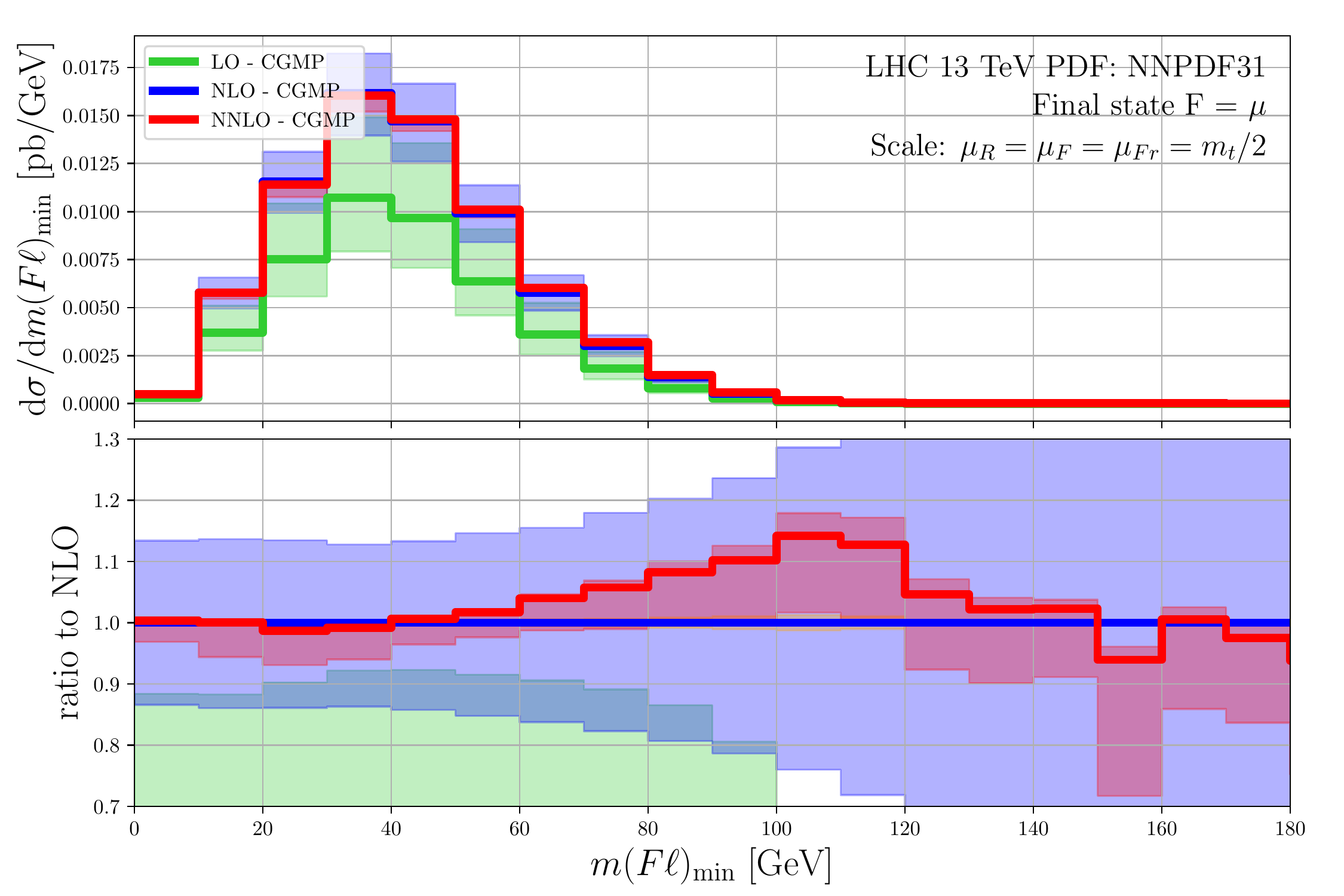}
\caption{The invariant mass $m(F\ell)_{\rm min}$ at LO (green), NLO (blue) and NNLO (red). Shown are the results for $F=B$ (left), $F=J/\psi$ (centre) and $F=\mu$ (right). The green, blue and red bands correspond to the scale uncertainty bands, while the yellow bands indicate the fragmentation function uncertainty.\vspace*{-4pt}}
\label{fig:mFl}
\end{figure}

Ref.~\cite{Czakon:2022pyz} studied this observable, also substituting the $B$-hadron for a $J/\psi$ meson or a muon to obtain analogous observables. The results are shown in fig.~\ref{fig:mFl}. Unlike fig.~\ref{fig:jetRatio}, these results use the new fragmentation functions fitted in ref.~\cite{Czakon:2022pyz}. Again, including the NLO corrections significantly reduces the scale uncertainties. This time, however, including the NNLO corrections leads to a much more dramatic reduction of the uncertainties, which reach the level of a few percent. Also shown in yellow is the fragmentation function uncertainty. However, this uncertainty is much smaller than the scale uncertainties and is only barely visible in fig.~\ref{fig:mFl}.\vspace*{-4pt}

\section{Conclusion}
I have presented some of the first results of a new framework for the computation of NNLO cross sections involving final-state hadrons. The results show massively reduced uncertainties compared to NLO, potentially improving the precision of future top-quark-mass measurements. Additionally, I have demonstrated that LHC data could be included in future $B$-hadron fragmentation function fits, potentially significantly improving their precision.\vspace*{-5pt}

\Acknowledgements
The work of T.G.~was supported by the DFG under grant 400140256 - GRK 2497: The physics of the heaviest particles at the Large Hadron Collider.

\end{document}